\documentstyle[epsfig]{mn}

\newif\ifAMStwofonts

\ifoldfss
  \ifCUPmtlplainloaded \else
    \NewTextAlphabet{textbfit} {cmbxti10} {}
    \NewTextAlphabet{textbfss} {cmssbx10} {}
    \NewMathAlphabet{mathbfit} {cmbxti10} {} 
    \NewMathAlphabet{mathbfss} {cmssbx10} {} 
  \fi
  \ifAMStwofonts
    \ifCUPmtlplainloaded \else
      \NewSymbolFont{upmath} {eurm10}
      \NewSymbolFont{AMSa} {msam10}
      \NewMathSymbol{\upi}     {0}{upmath}{19}
      \NewMathSymbol{\umu}     {0}{upmath}{16}
      \NewMathSymbol{\upartial}{0}{upmath}{40}
      \NewMathSymbol{\leqslant}{3}{AMSa}{36}
      \NewMathSymbol{\geqslant}{3}{AMSa}{3E}

    \fi
  \fi
\fi 

\ifnfssone
  \newmathalphabet{\mathit}
  \addtoversion{normal}{\mathit}{cmr}{m}{it}
  \addtoversion{bold}{\mathit}{cmr}{bx}{it}
  \newmathalphabet{\mathbfit} 
  \addtoversion{normal}{\mathbfit}{cmr}{bx}{it}
  \addtoversion{bold}{\mathbfit}{cmr}{bx}{it}
  \newmathalphabet{\mathbfss} 
  \addtoversion{normal}{\mathbfss}{cmss}{bx}{n}
  \addtoversion{bold}{\mathbfss}{cmss}{bx}{n}
  \ifAMStwofonts
    \ifCUPmtlplainloaded \else
      \UseAMStwoboldmath
      \makeatletter
      \new@mathgroup\upmath@group
      \define@mathgroup\mv@normal\upmath@group{eur}{m}{n}
      \define@mathgroup\mv@bold\upmath@group{eur}{b}{n}
      \edef\UPM{\hexnumber\upmath@group}
      \new@mathgroup\amsa@group
      \define@mathgroup\mv@normal\amsa@group{msa}{m}{n}
      \define@mathgroup\mv@bold\amsa@group{msa}{m}{n}
      \edef\AMSa{\hexnumber\amsa@group}
      \makeatother
      \mathchardef\upi="0\UPM19
      \mathchardef\umu="0\UPM16
      \mathchardef\upartial="0\UPM40
      \mathchardef\leqslant="3\AMSa36
      \mathchardef\geqslant="3\AMSa3E
    \fi
  \fi
\fi 

\ifnfsstwo
  \DeclareMathAlphabet{\mathbfit}{OT1}{cmr}{bx}{it}
  \SetMathAlphabet\mathbfit{bold}{OT1}{cmr}{bx}{it}
  \DeclareMathAlphabet{\mathbfss}{OT1}{cmss}{bx}{n}
  \SetMathAlphabet\mathbfss{bold}{OT1}{cmss}{bx}{n}
  \ifAMStwofonts
    \ifCUPmtlplainloaded \else
      \DeclareSymbolFont{UPM}{U}{eur}{m}{n}
      \SetSymbolFont{UPM}{bold}{U}{eur}{b}{n}
      \DeclareSymbolFont{AMSa}{U}{msa}{m}{n}
      \DeclareMathSymbol{\upi}{0}{UPM}{"19}
      \DeclareMathSymbol{\umu}{0}{UPM}{"16}
      \DeclareMathSymbol{\upartial}{0}{UPM}{"40}
      \DeclareMathSymbol{\leqslant}{3}{AMSa}{"36}
      \DeclareMathSymbol{\geqslant}{3}{AMSa}{"3E}
    \fi
  \fi
\fi 

\ifCUPmtlplainloaded \else
  \ifAMStwofonts \else 
    \def\upi{\pi}
    \def\umu{\mu}
    \def\upartial{\partial}
  \fi
\fi

\title[SCUBA Observations of NGC 4374]{SCUBA Observations of 
the Elliptical Galaxy NGC 4374}
\author[L. L. Leeuw et al. ]
       {L. L. Leeuw,$^{1,2}$\thanks{Email: leeuw@jach.hawaii.edu} A.E. Sansom,$^1$ E.I. Robson$^{2,1}$ \\
  $^1$Centre for Astrophysics, University of Central Lancashire,
Preston PR1 2HE \\
  $^2$Joint Astronomy Centre, 660 N. A'ohoku Place, Hilo, Hawaii 96720, USA}
\date{}

\pagerange{\pageref{firstpage}--\pageref{lastpage}}
\pubyear{1999}

\begin{document}

\maketitle

\label{firstpage}

\begin{abstract}

We present SCUBA imaging and photometry of the elliptical galaxy
NGC 4374.  The imaging observations are used to examine the spatial
distribution of thermal emission from dust and the 
radio to infrared continuum 
spectrum.  In the SCUBA 850\,$\mu {\rm m}$ image, the
galaxy is found to be a point
source, constraining the emission to less than $15''$ (1.5\,kpc for a
distance of 20.74 Mpc).  The simplest interpretation is that the
2000\,$\mu$m to 850\,$\mu$m SCUBA emission is synchroton from a
compact core or inner-jet, and most unlikely thermal emission from cold diffuse
dust.  We cannot exclude free-free emission, but this would be very
unexpected.  The thermal emission from dust is shown in $IRAS$ data
and these along with the 450\,$\mu$m SCUBA datum give a dust
temperature of 35\,K, corresponding to a
dust mass of $1.2 \times 10^{5} {\rm M}_{\odot}$.  These are pilot
observations in a program to look for cold,
diffusely distributed dust in elliptical galaxies, however much deeper
450\,$\mu$m imaging is required to investigate this in NGC 4374.    
\end{abstract}


\section{Introduction}

The interstellar medium (ISM) in elliptical galaxies is not easily 
probed by optical observations, however observations at a range of other 
wavelengths have revealed unexpected amounts of gas in these galaxies 
(Roberts et al. 1991). With the gas we also expect some dust to be 
present, since red giant stars lose dust to the ISM. Small amounts 
of dust are seen in dust lanes and patches in ellipticals (Sparks et 
al. 1985, Goudfrooij et al. 1994). The submillimetre (submm) 
camera SCUBA (Holland et al. 1999) provides 
an opportunity to test if there is cold (less than about $30$\,K), 
diffusely distributed 
dust present in elliptical galaxies as suggested from far-infrared 
observations.  Dust temperatures (25 to 35 K) and masses ($\sim10^4$ to 
few  $ \times 10^6 {\rm M}_{\odot}$) were estimated from $IRAS$ 
observations for a sample 
of ellipticals by Goudfrooij \& de Jong (1995). They showed that 
dust masses derived from $IRAS$ fluxes exceeded (by typically an order of
magnitude) the dust masses estimated from optically identified dust 
lanes and patches in many ellipticals.

However, the $IRAS$ data are limited for studying dust since $IRAS$
gave little information about the spatial distribution of the dust in
galaxies, so cannot be used to check directly if the dust is 
diffusely distributed or not. We also note that many ellipticals 
contain haloes of hot gas at around $10^7$\,K with masses of a 
few $\times 10^9 {\rm M}_{\odot}$ (Canizares, Fabbiano \& Trinchieri 1987). 
This X-ray emitting plasma is expected to destroy any dust grains
through sputtering by hot gas particles in a short time ($<10^7$ years, 
Draine \& Salpeter 1979). So, for diffusely distributed dust to be present 
in a typical giant elliptical galaxy, the dust would have to be protected 
or shielded from the plasma in some way. SCUBA, being an imaging device,
offers an opportunity to map out the distribution of any cool dust.

The giant elliptical NGC 4374 was chosen from the sample of Goudfrooij \&
de Jong (1995) as a candidate in which to look for diffusely distributed 
dust because it has optical dust lanes in the central region ($ <10''$).  
The dust mass estimated 
from the dust lanes is $3.5\times 10^4 {\rm M}_{\odot}$ and the dust mass 
estimated from the infrared ($IRAS$) fluxes (at 60 and 100\,$\mu {\rm m}$) is 
$1.35\times 10^5 {\rm M}_{\odot}$ (for a dust temperature of 35\,K), 
a factor of $\sim 4$ greater 
than from the optically identified dust (Goudfrooij \& de Jong 1995).
\footnote{Note that there was a typographical error in Goudfrooij \& 
de Jong 1995: the dust mass they estimated from the $IRAS$ data should have 
read Log$M_d$=5.13 (not 5.3) (Goudfrooij - private communication). 
We quote the correct dust mass here.} Since most of this 
dust is not seen in the optical dust lanes, Goudfrooij \& de Jong (1995) 
suggest that it must be diffusely distributed throughout the galaxy. 
Such dust would affect the colours in galaxies. However, this affect could 
be difficult to disentangle from age and metallicity variations in the 
stellar population, which also produce colour changes (Worthey 1994). 
Therefore the presence of a few million solar masses of dust, distributed 
throughout a galaxy, could well have escaped optical detection. On the other 
hand, spatially resolved submm observations, sensitive to thermal emission 
from cool dust, may directly reveal the presence of any such distribution. 

NGC 4374 (M84) is an E1 galaxy in the Virgo cluster lying at a distance
of 20.74\,Mpc.  It is classified as a steep spectrum Fanaroff-Riley I 
(FRI) radio galaxy with an infrared excess.  Two-sided jets
emerge from its compact core, which is less than 2 arcseconds
(Jenkins, Pooley \& Riley 1977).  In the context of the Blandford \&
Rees (1974) jet model, the jets transport energy from the core to
symmetrical, edge-darkened lobes which extend to about 2 arcminutes
and dominate the core emission (Laing \& Bridle 1987, Wrobel 1991).
The jets are perpendicular to the dust lanes.  There is a low ionization 
region near
the nucleus, extending for about 20 arcsecs along the direction of the
optical dust lanes (Baum, Heckman \& Breugel 1992). Bower et al.\ (1997) 
publish $HST$ images showing the central structure and
estimate a dust mass of $10^5 {\rm M}_{\odot}$ (for $D=20.74$ Mpc)
from their (V-I) image. 
The $HST$ images indicate central line emission from three components: 
a nuclear disk, an `ionization cone' and outer filaments.  

NGC 4374 
also has a hot X-ray halo with $\sim 10^9 {\rm M}_{\odot}$ of hot gas 
(Goudfrooij 1994). Its spectrum 
from the radio to infrared was previously studied by Knapp \& Patten (1991), 
who did not detect any submm emission above that expected from the 
continuation of the power-law emission at radio wavelengths. They used 
a previous generation detector in the submm (Duncan et al. 1990).   
Knapp \& Patten 
(1991) infer dust contents in a sample of radio galaxies (mostly 
ellipticals) of the order of that found in luminous spiral galaxies 
($\sim 10^6$ to $10^8 {\rm M}_{\odot}$).  They assumed a dust temperature of
18\,K ($\sim$ 15 to 20\,K estimated range) for the galaxies in their 
sample and estimated a dust mass of $2\times 10^6 {\rm M}_{\odot}$ in 
NGC 4374. Their low estimate of the dust temperature leads to a 
significantly larger dust mass estimate
than that of Goudfrooij \& de Jong (1995). 

In this paper we explore 
whether SCUBA jiggle mapping at 850 and 450\,$\mu$m and photometry at
2000, 1350, 850 and 450\,$\mu$m can confirm the presence of diffusely
distributed, cold dust in the giant elliptical galaxy NGC 4374.  
We use the SCUBA observations together with radio and infrared 
data from the literature to place tighter constraints on the spectral 
components, including the dust temperature and the size of the infrared
emitting region.

\section{Observations and Reductions}

\begin{table*}
 \centering
 \begin{minipage}{140mm}
  \caption{Submillimetre fluxes (Janskys) for NGC 4374 from SCUBA}
  \begin{tabular}{lllllllll}
UT Date	    &2000\,$\mu$m	    &1350\,$\mu$m	   &850\,$\mu$m   &450\,$\mu$m 
& 850/450\,$\mu$m      & 850/450\,$\mu$m  &
 $\tau_{\rm CSO}$ & Calibrators \\
           &146\,GHz	    &221\,GHz	   &350\,GHz   &677\,GHz 
      & 350/677\,GHz & 677/350\,GHz  & &  \\
&&&& &Total Int.  & Observing & & \\ 
&&&& & Time (s) & Mode & & \\ 
&&&& & &  & & \\ 
'990319  &$0.15\pm0.02$ &              &$0.18\pm0.02$ &$0.11\pm0.02$ &
 2160 & Photometry & 0.06 & Mars \\ 
'980214  &$0.15\pm0.02$ &$0.15\pm0.02$ &$0.11\pm0.02$ &$< 0.15$ &
 1260 & Photometry & 0.05 & CRL2688 \\
'980201  &              &              &$0.15\pm0.02$  &$< 0.12 $ &
 2640 & Mapping & 0.05 & IRC+10216 \\
'980122  &              &$0.14\pm0.02$ &$0.13\pm0.02$  &$< 0.12 $ &
 2340 & Mapping & 0.04 & IRC+10216 \\
'970703  &              &              &$0.16\pm0.03$        &   &
 2220 & Mapping & 0.1 & Uranus \\
\end{tabular}
\end{minipage}
\end{table*}

Imaging and photometric observations of NGC 4374 were obtained on
1997 September 3 as well as 1998 January 22, February 1, 
February 14 and 1999 March 19, with SCUBA, the Submillimetre Common-User
Bolometer Array on the JCMT (see Table 1).  In the imaging observations, we
operated the 91 bolometers of the short wave array (SW) at 450\,$\mu$m
and the 37 bolometers of the long wave array (LW) at 850\,$\mu$m,
giving beam widths of $8.5''$ and $14.5''$ (FWHM) respectively.  The
arrays have a 2.3\,arcminute field of view, and a dichroic
beam-splitter allows both arrays to be used simultaneously.  The
observations employed a 64-point jiggle pattern, fully sampling both
arrays.  Similarly, the submm photometric observations were obtained by 
operating
the central bolometers of the SW and LW arrays at $450\mu$m and $850\mu$m 
simultaneously, employing a 9-point jiggle pattern in a 3 by 
3 grid of $2''$.  The 1350\,$\mu$m and 2000\,$\mu$m observations used the
single photometry bolometers and also
employed a 9-point jiggle pattern.  Averaging the source signal in an 
area slightly
larger than the beam is intended to achieve the best photometry
accuracy under good-to-moderate seeing, and, in case of the
simultaneous observations with the LW and SW arrays, to also compensate 
for the very small offset between the arrays.   During the observations, the 
telescope was nodded and the secondary chopped in a specified scheme 
in order to eliminate sky emission as is convention in submillimetre 
and infrared astronomy.  

Residual sky emission was removed off-line 
by using quiet SCUBA array bolometers in which there
was no source emission, usually the bolometers in the first ring (for LW)
and second ring (for SW) from the centre (Jenness et al. 1998, Robson
et al. 1998). 
The pointing stability was checked before and 
after each map and before each photometry observation.  The focus was
checked every three hours or when there were noticeable dome temperature
fluctuations.  The atmospheric opacity was measured regularly by
performing SCUBA skydips.  The sky 
monitor at 225\,GHz on the Caltech Submillimetre Observatory (commonly
known as CSO tau or $\tau_{\rm CSO}$, see Masson 1993) updates every 
15 minutes and so was
used to monitor sudden changes in atmospheric opacity, and, in
cases where the SCUBA skydips were not measured or produced poor fits,
the CSO data were used to
extrapolate SCUBA atmospheric opacities as described in the SCUBA
observing manual.
  
The dedicated SCUBA data reduction software SURF (Jenness et al. 1998),
as well as the Starlink packages Kappa and Figaro were 
used to reduce and analyse the observations.  The data reduction 
consisted of first flat-fielding,  correcting for atmospheric extinction, 
and then residual sky emission removal (see above).   Noisy bolometers and 
integrations were blanked and spikes removed.  
For the 1350\,$\mu$m and 2000\,$\mu$m photometry observations no
residual sky emission removal was done as the observation used only 
a single bolometer.  Spike removal was performed by clipping the data at a
specified sigma.  The resulting data were
calibrated using instrumental gains that were determined from beam
maps of Mars or Uranus nightly and in the same observation mode as the
target observation  (see Table 1).   Planetary fluxes for each
filter were obtained using the JCMT utility program FLUXES.
During the January and February 1998 runs, the planets were not
available and the JCMT secondary calibrators CRL2688 and IRC+10216
were used 
(Sandell 1994; Sandell in preparation).  On nights when the $\tau_{\rm
CSO} > 0.07$ (see above and Table 1) the atmosphere was opaque to
450$\mu {\rm m}$ emission and observations could be made only at
longer wavelengths.
 
\section{Results and Discussion}

\begin{figure*}
\begin{centering}
\resizebox{5.0in}{!}{\includegraphics*[2mm,46mm][182mm,216mm]{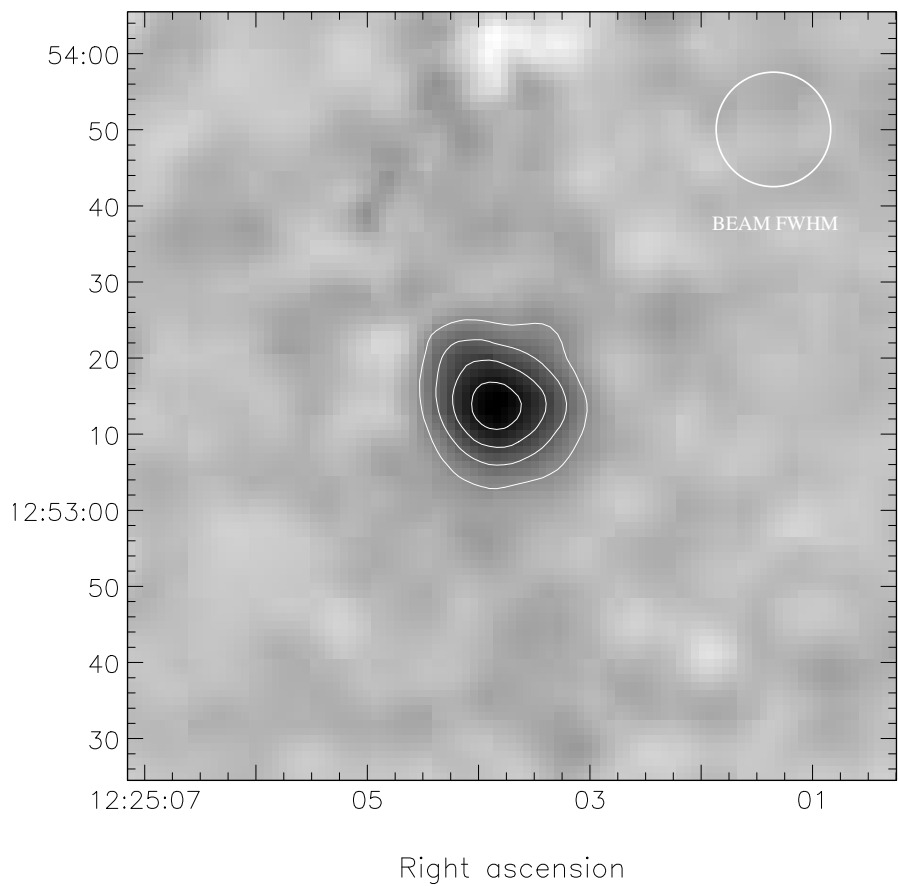}}
\caption{The SCUBA 850\,$\mu$m image of NGC 4374.  The observation was
made on 1998 February 1 when $\tau_{\rm CSO}$ was $\sim 0.05$.}
    \label{fig:n4374}
\end{centering}
\end{figure*}

Figure 1 shows the 850\,$\mu {\rm m}$ image of NGC 4374 together 
with the beam-width determined from observations of 3C 273, a standard 
point source for the JCMT, made in the same way and conditions as NGC 4374.  
The rms uncertainty on the map is less than 20\,mJy/beam.  The image of 
NGC 4374 has small northeastern and southwestern spurs that are not real 
features of the galaxy but a manifestation of the
beam smearing as a result of chopping.  These spurs
are in the chop direction and are detected at a much lower
level than the central emission.  An examination of the point spread 
function of the NGC 4374 image yielded FWHM$_{\rm minor-axis} = 15.0''$ 
and FWHM$_{\rm major-axis} = 15.7''$  The
result shows that the image of NGC 4374 is no more than $1''$ greater
than the JCMT beam.  This insignificant spatial extension is probably 
a result of a small pointing drift during the
long observation ($\sim 1$hr).  We conclude that NGC 4374 is not
extended in the SCUBA 850$\mu {\rm m}$ image which constrains the spatial
origin of the observed emission to less than $15''$ (1.5\,kpc for
a distance of 20.74\,Mpc) in diameter.  It is
worth noting that although the SCUBA beam is $\sim 15''$ at
850\,$\mu$m, the arrays have a 2.3\,arcminute field
of view, and therefore, at the distance of NGC 4374, we do not detect
any diffused emission to $\sim 13$\,kpc in extent, as we observed with
the array pointed at the centre of the galaxy and our observations
were spatially sensitive to the entire field of view of SCUBA.  
 
The results from the imaging and photometric observations are 
shown in Table 1. The uncertainties are a quadratic sum of 
the uncertainty arising from the measured signal-to-noise-ratio and a
systematic calibration uncertainty, which varies from 10\% at
850\,$\mu{\rm m}$ and lower, to 27\% at  450\,$\mu{\rm m}$.  The
450\,$\mu{\rm m}$ fluxes listed for the 1998 Feb 01 and 14 and Jan 22 runs
are $3\sigma$ upper limits.  The submillimetre flux variation over 
the four observing runs was within the errors on the fluxes. 

\subsection{Decomposing the SED}

\begin{figure*}
 \begin{minipage}{140mm}
    \leavevmode
      \epsfig{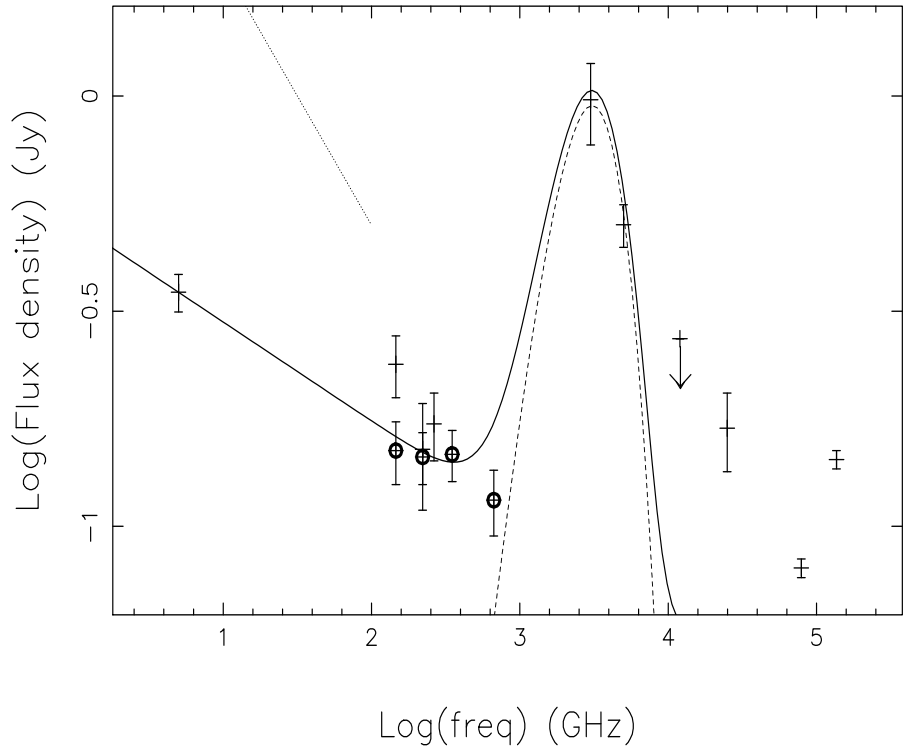}
\caption{Spectral energy distribution from radio to infrared for the
core of the elliptical galaxy NGC 4374. Filled circles are our data from SCUBA 
observations and the sources of other data are described in the text.
The solid line is a composite model of a power-law ($\alpha = -0.23$) 
plus (35\,K) greybody. This model fits the data quite well for the radio
through to 60$\mu$m infrared detections. The dashed line shows the 35K
greybody alone, showing that the SCUBA data lie above
this component. The dotted line is a power law of slope $- 0.6$, passing
through the integrated radio fluxes to 31\,GHz. The downward arrow 
indicates a 3 sigma upper limit at 25$\mu$m.}
    \label{fig:N4374_SED}
 \end{minipage}
\end{figure*}

Figure 2 shows the spectral energy distribution (SED) for NGC 4374,
including data from radio (Jenkins, Pooley \& Riley 1977) to infrared ($IRAS$)
wavelengths. The infrared fluxes 
were obtained from NED (the NASA/IPAC Extragalactic Database) and have
been interpreted as thermal emission from dust (Knapp
\& Patten 1991, Goudfrooij 1994).  At
longer radio wavelengths, the lobes dominate the integrated flux from
NGC 4374 (by a factor of 8 at 5\,GHz, Wrobel 1991) and give an
excellent power law of spectral index (${\rm S}_{\nu} \propto
\nu^{ \alpha} $) of $\alpha_{31\,{\rm
GHz}}^{2.7\,{\rm GHz}} = - 0.6 \pm 0.03$, as shown by the dotted line
in Figure 2.  This spectral index is consistent with the
classification of NGC 4374 as a steep spectrum radio galaxy.  The good fit
suggests variability larger than the errors on the fluxes 
is uncommon for the synchrotron power-law of the radio emission.

The submm data points for fluxes within the SCUBA beam and previous
submm measurements
fall below an extrapolation of this steep radio power-law from the
integrated flux.  We see no sign of extended emission from the radio
lobes on any of our maps, and the $3\sigma$ upper
limit on the 850\,$\mu$m integrated flux within a region approximately the
size of the lobes is 0.6\,Jy.  This is much higher than the extrapolation of the radio
power-law from the integrated fluxes, showing that we are insensitive
to the radio lobes (Figure 1).  As we show later, this indicates that
if the observed 
SCUBA 850\,$\mu$m emission is non-thermal in nature it is dominated
by the compact core (or inner-jet).

\subsubsection{The Core SED}

In Figure 2 we have plotted the single available core radio flux 
for NGC 4374, at 5GHz (Jenkins, Pooley \& Riley 1977, Wrobel 1991); the 
circles are the means of the SCUBA data listed in Table 1.  
The 5\,GHz core flux together with the SCUBA data give a power-law 
slope of $\alpha = -0.23 \pm 0.03$.  In compact radio galaxies, such flat
spectra have been found to comprise spectra from many unresolved
components, each with its own synchrotron spectrum, the sum of which
results in the observed flat power-law.  Although we do not wish to
over-interpret our results, since we only have one radio core point and
the SCUBA data, we note that the observed flatter power-law of the
NGC 4374 core could be a consequence of such unresolved synchrotron
components.  At 5GHz, the northern jet in NGC 4374 is brighter than
the southern jet for the first 10 arcseconds (Bridle and Perley 1984).
This one-sided inner-jet  may comprise such unresolved components.

It is worth noting that the SCUBA fluxes at 850\,$\mu$m, $1350\,\mu$m and 
2000\,$\mu$m alone give a spectral index of $\alpha_{146\,{\rm
GHz}}^{350\,{\rm GHz}}  
= - 0.16 \pm 0.32$ in the submm, forming a flat spectrum resembling
that of a free-free emission component.  While this 2000\,$\mu$m to 
850\,$\mu$m spectral index is consistent with that of free-free emission 
($\alpha = -0.1$), the large uncertainty on its value limits our ability to
conclusively prove its origin.  Our
450\,$\mu$m point lies well below this $\alpha = -0.16$ spectrum, which does not
help to prove or disprove its origin, but instead adds the complication that
this free-free component might have a high-frequency cut-off between 850\,$\mu$m and 450\,$\mu$m.  In steep spectrum radio galaxies 
such as NGC 4374,
it is uncharacteristic for a free-free component to dominate the
power-law in the mm-submm wavelength range.  Therefore if this component is
real it would have to come from very high ionization and it would be
an interesting discovery.  We cross-correlated the `submm flat spectrum 
component' in Figure 2 with the well known spectrum of
3C273 using data taken with SCUBA on 1998 February 15, just one
night after and under very
similar conditions as the photometry
observations of NGC 4374.  The spectrum of 3C273 gives a much steeper
slope and a spectral index of $\alpha_{146\,{\rm
GHz}}^{350\,{\rm GHz}} = - 0.7 \pm 0.1$, as expected of this variable
object during its quiescence phase (Robson et al. 1993).  Therefore with our
current, limited data we cannot say the 850\,$\mu$m, $1350\,\mu$m and 
2000\,$\mu$m fluxes come from significant
free-free emission, if any at all.
  
To model the radio to infrared SED we used a combination of a power-law 
(non-thermal radio
emission from the active nucleus) plus greybody (re-processed emission
from dust in the infrared, Hughes, Gear \& Robson 1994). The greybody
allows for the fact that we see to different physical depths at different
wavelengths, for a given optical depth. The composite model flux is then:

\begin{equation}
F_{\nu} = C\nu^{-\alpha} + \Omega B_{\nu}(T)
[1-{\rm exp}(-({\frac{\lambda_o}{\lambda}})^{\beta})].
\end{equation}
In equation (1) $C$ is the normalisation for the power-law component,
$\alpha$ the power-law spectral slope, $\Omega$ the solid angle for
the greybody component,  $B_{\nu}(T)$ the Planck function at
temperature $T$, $\lambda_o$ the wavelength at which the optical depth
is unity and $\beta$ the emissivity index of the grains.  The 
normalisation factors ($C$ and $\Omega$) are obtained by forcing the
model to agree with the 5GHz core radio and IR (60 and 100\,$\mu$m) fluxes
respectively, since the two components dominate in these different 
wavebands. The power-law index and its uncertainty were estimated 
from linear regression fits to the core radio plus SCUBA data in a 
log-log plot.  We assume that the $IRAS$ emission comes from a region less
than $\sim 10''$ due to lack of extension seen at 450\,$\mu$m. 
The best fit temperature and its uncertainty for the greybody component 
were estimated by eye from
plots of the composite model versus the data. Other parameter values
($\lambda_o = 7.9\,\mu$m and $\beta = 1.3 \sim\pm 0.5$) were fixed at
the values given in Hughes et al.\ (1994) in their study of M82.
The parameter values determined from the model are summarised in
Table 2, and the model is plotted as a solid line in Figure 2.

\begin{table}
 \centering
  \caption{Parameters determined from the model fit to the SED for NGC 4374}
  \begin{tabular}{llll}
Parameter &Value    &Error    &Units  \\
&&& \\
$\alpha$     &-0.23 &$\pm 0.03$ & - \\
$\Omega$     &3.96E-11 & - &Steradians \\
 & 1.46 & - & arcseconds \\
$T$          &35 &$\pm 5$ &K \\
\end{tabular}
\end{table}

We note that the greybody fit demands that the warm dust emission
comes from a very compact zone, much smaller that the SCUBA
450\,$\mu$m resolution (see Table 2).  Because the 450\,$\mu$m data
point lies below the sum of the thermal and non-thermal emission (see
Figure 2), the simplest explanation is that the synchrotron spectrum
steepens between 850\,$\mu$m and 450\,$\mu$m (or somewhat long-ward of
850\,$\mu$m, given the uncertainties).  Also, some small part of the 
450\,$\mu$m emission could be due to low surface brightness dust of
$\sim 15\,$K to 20\,K.  Better spectral 
coverage and higher signal-to-noise data in the submm are clearly needed to
test more complex spectral models.

\subsection{Dust Mass and its Implications}

The temperature that fits the $IRAS$ data and is constrained 
by the SCUBA 450\,$\mu$m measurement is  $\sim 35 \pm 5$\,K (see Figure 2). 
Not surprisingly, this is the same temperature as Goudfrooij (1994) 
found for this galaxy from his analysis of the $IRAS$ data alone. Changing
$\beta$ between 2.0 and 1.0 makes very little difference to the best
fit temperature.   
The mass of emitting dust $M_{\rm d}$ can  be derived from a simple model
adapted from Hildebrand (1983), where
  
\begin{equation}
 M_{\rm d} = \frac{S_{\nu} D^2}{k_{\rm d} B_{\nu}(T)}.
\end{equation}
In equation (2) $S_{\nu}$ is the measured flux, $D$ the distance to 
the source (20.74\,Mpc for NGC 4374 as assumed by Goudfrooij 1994),
$B_{\nu}(T)$ the Planck function and $k_{\rm d}$ the grain mass absorption
coefficient.  We assumed $k_{\rm d}^{100\,\mu {\rm m}} = 2.5\,{\rm m}^2 
{\rm kg}^{-1}$ (Hildebrand 1983) and estimated a dust mass of 
$1.2 \times 10^5 {\rm M}_{\odot}$ for ${S_{100\,\mu{\rm m}}}=0.98 
\pm 0.21$\,Jy and $T =35$\,K.   The dust mass we have calculated for 
NGC 4374 is about the same as Goudfrooij \& de Jong (1995), as
expected since they used the same temperature.  Furthermore, it is
similar to that found in a recent $HST$ ($V-I$) study of dust lane in NGC
4374 by Bower et al. (1997), who found the dust mass to be in agreement 
with that derived from $IRAS$ data, 
in contrast with the dust {\em deficit} found in other optical studies 
of ellipticals (Goudfrooij \& de Jong 1995).  However for NGC 4374 it is at 
least one order of magnitude lower than Knapp \& Patten (1991), who 
assumed a cooler temperature 
of 18\,K (based on objects in their sample of nearby radio galaxies for 
which $1350\,\mu$m, 800\,$\mu$m and 450\,$\mu$m fluxes were detected).
Also, they obtained the higher mass even though they assumed a smaller 
distance of 13\,Mpc.  While the low dust temperatures estimated by Knapp \&
Patten (15 to 20 K) are similar to the Galactic Plane value of $T=19$\,K,
they are inconsistent with the SCUBA plus $IRAS$ observations of NGC 4374
if the dust is assumed to be all a single temperature.

As for the extended radio lobe emission, we have not detected
the extended low-level emission from diffusely distributed dust with
SCUBA.  We have shown that the 2000\,$\mu$m to 850\,$\mu$m data are
most unlikely due to diffuse cold dust, but some small part of the
450\,$\mu$m emission could be due to dust of $\sim 15\,$K to 20\,K.
 Also, our $3\sigma$ upper 
limit to the surface brightness at 450\,$\mu$m is too high and not
useful for setting dust mass upper limits on this possibly colder dust.
Deeper SCUBA imaging observations are clearly needed to address this
question and will be undertaken in the 2000 observing season.

\section{Conclusions}

Following the suggestion that elliptical galaxies may contain diffusely
distributed dust (Goudfrooij 1994) we searched for this dust with
submm imaging observations of the elliptical galaxy NGC 4374, using SCUBA 
on the JCMT.  We have not detected
low-level, diffusely distributed dust with SCUBA.  The emission at
850\,$\mu$m is spatially unresolved (diameter $< 15''$; 1.5\,kpc). 

Adding the SCUBA submm data to existing radio through IR data for this 
galaxy we can constrain the dust component to a single temperature of 
$\sim$ 30 to 40\,K, implying a dust mass of
$\sim 1.2\times 10^5 {\rm M}_{\odot}$.  The model fitting in section
3.1 gives an 
angular extent of the dust to be $\sim 1.5''$ in diameter.  Mindful
that this result is too simplistic for a radiative model of an AGN
torus, we note in passing that if we assume a co-mixing of the
molecular gas and dust, this constrains the size of any molecular
torus around the AGN 
core of NGC 4374 to $\sim 150$\,pc in diameter.  

The mm-submm observations show a flat spectral index
that is consistent with the 5\,GHz radio core flux.  The spectral
index of the 850\,$\mu$m
to 2000\,$\mu$m fluxes alone is consistent with that
expected from free-free emission. The
possibility of a free-free component is very unusual and intriguing.
Given the care we took over calibration, we see no systematic reasons
for our peculiar mm-submm SED.  Future observations of NGC 4374 will
aim to achieve very high signal-to-noise observations in order to
determine the spectral index to a much higher accuracy and therefore
prove or disprove the presence of a third spectral component.

In future observations of dusty ellipticals, we will
concentrate on those with luminous infrared emission, coldest dust and
less non-thermal radio emission
in order to try to resolve the distribution of cold dust in
ellipticals.  We will obtain short mm images of the
central regions of NGC 4374 to determine the spatial extent of the
emitting region(s) and deeper SCUBA imaging observations to address
the possibility of extended low-level emission from cold dust.  These observations together 
with $ISO$ data will be used in order
to get a tighter constraint, and thus a better handle, of the radio to submm
SED.  These observations have important consequences for the interpretation
of colours and colour gradients in elliptical galaxies, which are
attributed to age and metallicity changes in the absence of diffusely
distributed dust. 

\section*{Acknowledgments}
Thanks to Rob Ivison for helping with some of the observations and to
Goran Sandell for useful discussion during the reduction and analysis
of the SCUBA images.  LLL acknowledges the University of Central Lancashire for
a full-time research studentship.  We thank an anonymous referee for
very helpful comments and for pointing out the possibility for cold
diffuse dust.  The JCMT is operated by the Joint 
Astronomy Centre, 
on behalf of the UK Particle Physics and Astronomy Research Council, 
the Netherlands Organization for Scientific Research and the Canadian 
National Research Council.  This research has made use of the NASA/IPAC 
Extragalactic Database (NED) which is operated by the Jet Propulsion 
Laboratory, California Institute of Technology, under contract with the 
National Aeronautics and Space Administration.

\label{lastpage}

\end{document}